# Single-atom anchored novel two-dimensional MoSi$_2$N$_4$ monolayers for efficient electroreduction of CO$_2$ to formic acid and methane

Wei Xun[a*], Xiao Yang[a], Qing-Song Jiang[a], Ming-Jun Wang[b*], Yin-Zhong Wu[c], Ping Li[d*]

a Faculty of Electronic Information Engineering, Huaiyin Institute of Technology, Huaian 223003, China (xunwei@hyit.edu.cn)

a Faculty of Electronic Information Engineering, Huaiyin Institute of Technology, Huaian 223003, China (yangxiao@hyit.edu.cn)

a Faculty of Electronic Information Engineering, Huaiyin Institute of Technology, Huaian 223003, China (jiangqingsong05@hyit.edu.cn)

b School of Automation and Information Engineering, Xi'an University of Technology, Xi'an, Shaanxi 710048, China (wangmingjun@xaut.edu.cn)

c School of Physical Science and Technology, Suzhou University of Science and Technology, Suzhou 215009, China (yzwu@usts.edu.cn)

d State Key Laboratory for Mechanical Behavior of Materials, Center for Spintronics and Quantum System, School of Materials Science and Engineering, Xi'an Jiaotong University, Xi'an, Shaanxi 710049, China (pli@xjtu.edu.cn)

**Abstract**

Efficient and selective CO$_2$ electroreduction into value-added chemicals and fuels emerged as a significant approach for CO$_2$ conversion, however, it relies on catalysts with controllable product selectivity and reaction paths. In this work, by means of first-principles calculations, we identify five catalysts (TM@MoSi$_2$N$_4$, TM = Sc, Ti, Fe, Co and Ni) comprising transition-metal atoms anchored on a MoSi$_2$N$_4$ monolayer, whose catalytic performance can be controlled by adjusting the *d*-band center and occupation of supported metal atoms. During $CO_2$ reduction, the single metal atoms function as the active sites activates the $MoSi_2N_4$ inert basal-plane, and as-designed electrocatalysts exhibit excellent activity in $CO_2$ reduction. Interestingly, HCOOH is the preferred product of $CO_2$ reduction on the Co@$MoSi_2N_4$ catalyst with a rate-determining barrier of 0.89 eV, while the other four catalysts prefer to reduce $CO_2$ to $CH_4$ with a rate-determining barrier of 0.81-1.24 eV. Moreover, $MoSi_2N_4$ is an extremely-air-stable material, which will facilitate its application in various environments. Our findings provide a promising candidate with high activity, catalysts for renewable energy technologies, and selectivity for experimental work.

## 1. INTRODUCTION

Because of the continuing increase in the emissions of $CO_2$ from excessive fossil fuel usage, the reduction of $CO_2$ into environmentally friendly, high-efficiency, and low-cost alternative fuels such as formic acid (HCOOH), methanol ($CH_3OH$), and methane ($CH_4$) is recognized as one of the most promising approaches that would positively impact the global carbon balance and energy storage[1-6]. Since $CO_2$ is an extremely stable and nonreactive molecule, converting $CO_2$ into fuels is a scientifically challenging problem requiring appropriate catalysts and high energy input[2]. Single atom catalysts (SACs), first proposed for CO oxidation in 2011[7], provide efficient activation and conversion of $CO_2$ using transition metal (TM) atoms[8–16]. The coexistence of empty and occupied TM *d*-orbitals can accept lone-pair electrons, and then back-donate these electrons to the antibonding orbitals to weaken the C=O bonds.

Remarkably, since two dimensional (2D) materials have large surface-to-volume ratios, short carrier diffusion distances, unique electronic properties, and abundant active sites, they serve as promising substrates for atomically dispersed transition metal atoms for $CO_2$ reduction. To date, most of the discovered monolayer 2D materials have a thickness of n ⩽ 7. Graphene monolayer is famous six-membered ring (SMR) materials[17] of n = 1, which have been a central topic for $CO_2$ reduction[18–28]. The silicon counterpart of graphene (n = 2), has also shown great potential for $CO_2$ reduction[29, 30]. For n = 3, monolayer transition-metal dichalcogenides are the most studied SMR materials. In particular, the reduction of $CO_2$ to methanol can be achieved with the $MoS_2$ supported single Co atom catalyst[31]. Among the family of n = 4, monolayer group III chalcogenides have attracted growing interest. For instance, TM@InSe catalysts based on 2D InSe and transition metal atoms are candidates for CO, HCOOH, and $CH_4$ production[32]. The 2D $In_2Se_3$, the representative systems of n = 5, Anchoring different single TM atoms shows the great electrocatalytic ability of $CO_2$ reduction via ferroelectric switching[33]. Recently, CaMg was found to have a sextuple layer (n = 6) structure, but there is no research about $CO_2$ reduction[34].

Recently, a new compound of septuple layer SMR material, $MoSi_2N_4$, has been successfully, which has a band-gap of ~1.94 eV with excellent ambient stability[35]. The $MoSi_2N_4$ monolayer can be built by intercalating a 2H-$MoS_2$-type $MoN_2$ layer (n = 3) into an α-InSe-type $Si_2N_2$ (n = 4). Motivated by the extensive research attention of $MoSi_2N_4$ and the reported interesting electronic and catalytic properties[36–46], it is of great fundamental interest to determine whether the emerging 2D $MoSi_2N_4$ materials can be applied to other important electrocatalytic reactions, such as the $CO_2$ reduction reaction ($CO_2$RR).

In this work, we theoretically investigate the potential of using transition metal decorated $MoSi_2N_4$ monolayers for electrochemical $CO_2$ reduction into hydrocarbon fuels. Our efforts identify five catalysts with singly dispersed TM atoms anchored on the 2D $MoSi_2N_4$ monolayer (denoted as TM@$MoSi_2N_4$, where TM = Sc, Ti, Fe, Co and Ni). The differential charge density demonstrated that TM atoms form strong interactions with $MoSi_2N_4$ by exchanging electron density. With the increase in atomic number, the net electron transfer of TM atoms to $CO_2$ decreases, indicating that $CO_2$ activation changes from strong to weak. Anchoring different TM atoms can not only alter the reaction barrier and paths of $CO_2$ reduction, but also lead to different final products. These performance improvements stem from the synergistic effects of the adjusted empty and occupied *d*-orbitals (*d* orbital center) of the adsorbed metal atom, electron transfer, and $CO_2$ adsorption energies. These SACs and catalytic mechanisms introduce a feasible approach to significantly improve the efficiency of the $CO_2RR$.

## 2. COMPUTATIONAL METHODS

DFT calculations were performed by using the Vienna Ab Initio Simulation Package (VASP)[47, 48]. The exchange−correlation interactions were treated within the generalized gradient approximation (GGA)[49] in the form of the Perdew−Burke−Ernzerhof (PBE) functional[50]. The van der Waals interactions were described using the Bayesian Error Estimation Exchange-correlation functional (BEEF-vdw)[51]. The electron wave functions were expanded using plane waves with a cutoff energy of 500 eV, and the convergence criteria for the residual force and energy on each atom during structure relaxation were set to 0.002 eV/Å and $10^{-6}$ eV, respectively. The vacuum space was more than 20 Å, which was enough to avoid interactions between periodic images. The dipole correction is taken into account for all the asymmetric structures[52]. The single-atom catalysts were modelled by depositing one metal atom on 2×2×1 supercell $MoSi_2N_4$. The Brillouin zone (BZ) was sampled with a Monkhorst−Pack mesh with a 6 × 6 × 1 kpoint grid in reciprocal space during geometry optimization, and a 12×12×1 kpoint grid to calculate the electronic properties of all systems. The Gibbs free energy is calculated using a hydrogen electrode model (CHE)[53], and the solvent effect is considered with the implicit solvent model implemented in VASPsol[33, 54]. The site-specific charge differences were obtained using Bader analysis.

## 3. RESULTS AND DISCUSSION

Due to the large band gap (1.94 eV) of the $MoSi_2N_4$ monolayer, not enough electrons are injected into the antibonding $2\pi_u$ orbitals of $CO_2$ so that the strong *sp*-hybridization symmetry of the carbon atom cannot be disrupted[55]. Therefore, the $MoSi_2N_4$ material itself is not suitable as a catalyst for $CO_2$ reduction. Our theoretical study also confirmed this point: by adsorption on $MoSi_2N_4$, the inherent linear O=C=O structure of $CO_2$ molecules can be well maintained (see Supplementary Fig. S1).

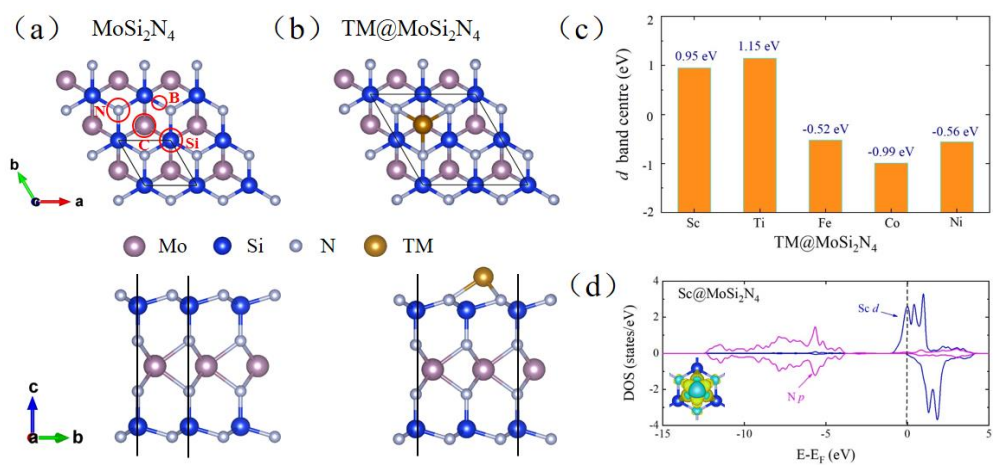

Fig. 1 Geometries and electronic structures of the MoSi$_2$N$_4$ catalysts. Top and side views of the optimized MoSi$_2$N$_4$ monolayer without transition metal atoms (a) and with transition metal atoms (b). Red circles denote selected adsorption sites: N, top of the N atom; B, top of the Si-N bond; Si, top of the Si atom; C, center of the six-membered ring. The black rhombus represents the unitcell of MoSi$_2$N$_4$ and supercell of TM@MoSi$_2$N$_4$. (c) The $d$ band center of TM@MoSi$_2$N$_4$ (TM =Sc, Ti, Fe, Co and Ni). (d) The partial density of states of Sc@MoSi$_2$N$_4$. The isosurfaces are 0.005 e/Å$^3$.

To activate CO$_2$ electrochemical reduction, we introduce transition metal atoms to modify the MoSi$_2$N$_4$ monolayer and provide additional electrons to break strong $sp$ hybridization, to activate CO$_2$ molecules. Compared with traditional metal catalysts, SACs usually show higher catalytic performance because of their high low-coordination configuration ratio. However, transition metals must be screened because only SACs with well empty/occupied $d$ orbitals balanced can show the best catalytic performance[7, 33, 56, 57]. To find suitable SACs, 10 transition metal atoms (3$d$) were selected for chemisorption on the monolayer surface of MoSi$_2$N$_4$ and the most favorable energy configuration was determined (see Supplementary Fig. S2). Then，we evaluated all composites against two key criteria. Firstly, in order to ensure the stability of the active center, a single transition metal atom should be stably adsorbed on the surface of MoSi$_2$N$_4$ monolayer without destroying the underlying structure. Second, a single transition metal atom should be able to activate CO$_2$, that is, the linear structure of O=C=O should decompose after adsorption. After comprehensively studying the adsorption configuration with the most favorable energy (Fig. 1a, 1b) and its stability, we selected five TM atoms (TM=Sc, Ti, Fe, Co and Ni) as promising $CO_2$ reduction catalysts for further study. The other five candidates are either unstable (TM=Mn, Zn) or not effective at activating $CO_2$ molecules (TM=V, Cr, Cu) (see Supplementary Fig. S3). To further demonstrate the structural stability of the SACs chosen in this study, ab initio molecular dynamics (AIMD) simulations were also performed at TM@$MoSi_2N_4$ with a Nose-Hoover thermostat at 300 K. In the AIMD simulations, the metal atoms were anchored in energetically favorable positions for at least 15 ps, even at room temperature (300 K) (see Supplementary Figure S4).

Table 1 Parameters for the pure TM@$MoSi_2N_4$ and $CO_2$ adsorbed TM@$MoSi_2N_4$: adsorption site ($S_{ad}$), binding energies of TM atom ($E_{b-TM}$ in eV/atom) and $CO_2$ molecule ($E_{b-CO2}$ in eV/molecule), charge lost from the adsorbed TM atoms ($Q_{TM}$ in e/atom), average TM-N bond length ($l_{TM-N}$ in Å), charge gained by the adsorbed $CO_2$ molecule ($Q_{CO2}$ in e/molecule), and bond angle (∠OCO in °).

| Catalysts | $S_{ad}$ | $E_{b-TM}$ | $Q_{TM}$ | $l_{Tm-N}$ | $E_{b-CO2}$ | $Q_{CO2}$ | ∠OCO |
|---|---|---|---|---|---|---|---|
| Sc@$MoSi_2N_4$ | C | -0.60 | 1.45 | 2.51 | -2.72 | 1.10 | 127.3 |
| Ti@$MoSi_2N_4$ | C | -0.81 | 1.38 | 2.43 | -2.11 | 1.05 | 129.5 |
| Fe@$MoSi_2N_4$ | C | -0.12 | 0.74 | 2.08 | -1.26 | 0.65 | 142.5 |
| Co@$MoSi_2N_4$ | C | -0.41 | 0.50 | 2.07 | -1.46 | 0.42 | 147.8 |
| Ni@$MoSi_2N_4$ | C | -1.17 | 0.46 | 2.06 | -1.52 | 0.45 | 147.4 |

At the most energetically favorable positions, all five TM atoms have strong binding energies (Table 1), indicating that the interaction between the metal atom and the base is strong enough, as seen from the strong TM-N bond (Fig. 1a and supplementary Fig. S5). Chemical bond interactions can be illustrated by the large accumulation of partial charges between TM atoms and surrounding N atoms (see Table 1 and supplementary Fig. S5) and the strong hybridization of N $p$ and TM $d$ orbitals

(see Fig. 1d and supplementary Fig. S6). The average TM-N bond length, binding energy and electron transfer are shown in Table 1. As the atomic number of metals increases, TM@MoSi$_2$N$_4$ usually has a shorter average TM-N bond length and less electron transfer. The balance between the empty and occupied $d$ orbitals caused by electron transfer has a significant impact on the TM@MoSi$_2$N$_4$ catalytic activity. With Sc@MoSi$_2$N$_4$ as an example, the charge is redistributed between the Sc atom and N atom (see Fig. 1d), and the $d$ orbital of the Sc atom is exhausted. Compared with Ni@MoSi$_2$N$_4$ (0.46 e, see supplementary Fig. S6), Sc@MoSi$_2$N$_4$ has more electron transfer (1.45 e), and the Sc-$d$ orbital moves to higher energies. As shown in Fig. 1c, the $d$-band center of Sc@MoSi$_2$N$_4$ (0.95 eV) is higher than that of Ni@MoSi$_2$N$_4$ (-0.56 eV), indicating that Sc@MoSi$_2$N$_4$ has a better catalytic capacity[53]. Under the same mechanism, the other four TM@MoSi$_2$N$_4$ catalysts show a similar phenomenon (see Fig. 1c): the $d$-band center of the TM atom moves in a lower energy direction as the metal atomic number increases. Note that TM@MoSi$_2$N$_4$ (TM = Sc, Ti) and TM@MoSi$_2$N$_4$ (TM = Fe, Co, Ni) have little difference in their $d$-band center positions. These comparisons show that both reaction paths and barriers are influenced by metal atoms (see detailed classification of catalysis below).

The effective activation of CO$_2$ molecules is the key to subsequent reduction. According to the structural analysis, all TM@MoSi$_2$N$_4$ catalysts can effectively activate CO$_2$ molecules by forming bidentate C-TM-O species (see Fig. 2 and supplementary Fig. S7). In addition, the binding energy of catalysts with active CO$_2$ molecules (-2.72 to -1.26 eV; see Table 1) is equivalent to the reported catalysts[28, 33, 50]. Strong hybridization of O-$p$ orbitals and TM-$d$ orbitals and significant charge transfer between TM@MoSi$_2$N$_4$ and CO$_2$ ensure that inert molecules are chemically captured (see Fig. 2a, 2b and Supplementary Fig. S7). Note that the activation degree of CO$_2$ can be clearly seen from the angles of ∠OCO (see Table 1). At the same time, the activation degree of CO$_2$ on TM@MoSi$_2$N$_4$ (TM = SC, Ti) was significantly higher than that on TM@MoSi$_2$N$_4$ (TM = Fe, Co, Ni), as shown by the greater binding energy (- 2.72 ~ -2.11 eV vs-1.26 ~ -1.52 eV), more electron transfer (1.05 ~ 1.1 e vs 0.42 ~ 0.65 e) and smaller ∠OCO angle (127.3º ~ 129.5º vs 142.5º ~ 147.8º).

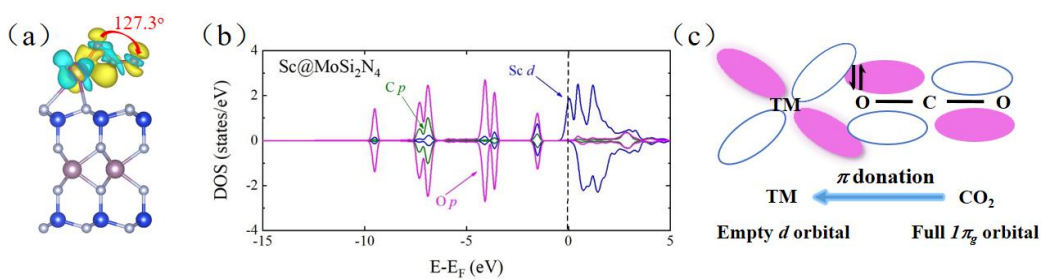

Fig. 2 Transition metal atom activated $CO_2$. (a) The differential charge density plots and (b) partial density of states of $CO_2$ adsorbed on Sc@MoSi$_2$N$_4$. (c) Simplified schematic diagrams of $CO_2$ bonding to transition metal atoms.

The diverse behavior of $CO_2$ on different surfaces is attributed to the catalytic activation mechanism of $CO_2$, i.e., the coexistence of charge dissipation and accumulation between TM atoms and $CO_2$ molecules. The empty TM $d$ orbital can accept electrons from the highest occupied molecular orbital ($1\pi_g$ orbital) in the $CO_2$ molecule (Fig. 2c). The synergistic effect of electron reception, feedback, and $d$-orbital occupation ensures that $CO_2$ can be activated efficiently. The isolated Sc atom has only one electron occupying the $d$ orbital ($3d^1$). After adsorption on the MoSi$_2$N$_4$ substrate, the Sc atom loses part of its $d$ electrons, forming a Sc-N bond, and the resulting empty Sc $d$ orbital provides a channel for electron acceptance and donation to activate the $CO_2$ molecule. Compared with other TM@MoSi$_2$N$_4$ catalysts, under the circumstance that the intrinsic part of the TM atom occupies the $d$ orbital, the Sc@MoSi$_2$N$_4$ catalyst has a higher degree of $CO_2$ activation, a relatively smaller angle ∠OCO and a larger charge transfer (see Table 1).

As an important competitive side reaction, hydrogen evolution (HER) may significantly inhibit the Faraday efficiency of the $CO_2$RR by depleting the proto-electron pairs in electrolyte solutions[58, 59]. To verify whether the $CO_2$RR is more favorable, we first calculated the change in Gibbs free energy (ΔG) in the first step of $CO_2$RR (*+$CO_2$+H$^+$+e$^-$ → OCOH* or OCHO*) and HER (*+H$^+$+e$^-$ → H*). According to Brønsted Evans Polanyi relations[60, 61], reactions with lower ΔG values have smaller reaction barriers and are therefore more dynamically advantageous.

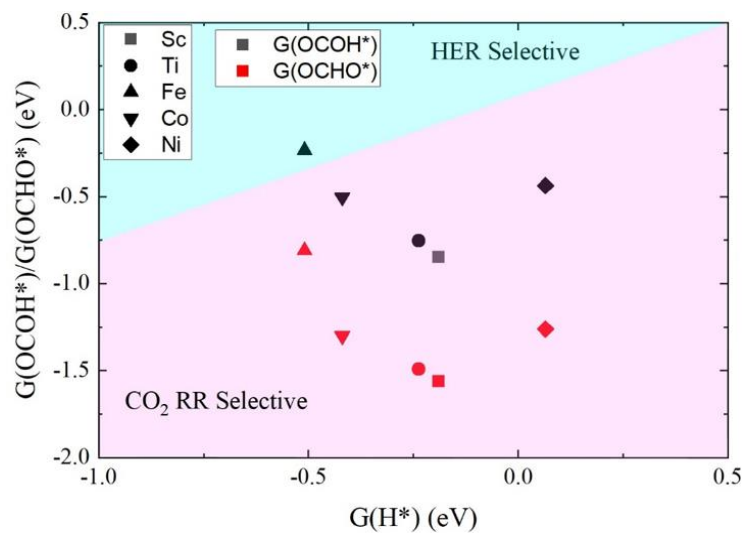

Fig. 3 Selectivity for CO$_2$RR vs HER. Gibbs free energy changes (ΔG) of the initial protonation of CO$_2$RR vs. HER on TM@MoSi$_2$N$_4$ (TM= Sc, Ti, Fe, Co and Ni). Data points in the purple region indicate higher selectivity toward the CO$_2$RR, while those in the blue region indicate higher selectivity toward the HER. Black and red symbols represent ΔG (OCOH*) vs ΔG (H*) and ΔG (OCHO*) vs ΔG (H*), respectively.

To a clearer understanding of the mechanism, we investigate the possible reaction pathways of the CO$_2$RR on TM@MoSi$_2$N$_4$, as shown in Figure 4. During CO$_2$RR, the reaction includes eight elemental hydrogenation steps, which are determined by the most stable product at each step. The initial step of CO$_2$RR is the formation of OCHO*, which is followed by the hydrogenation of OCHO* to give HCOOH*. (OCHO* + H$^+$ + e$^-$ → HCOOH*). Note that while the decomposition of HCOOH into CHO* + H$_2$O (I) is induced by the action of the third hydrogen, desorbed HCOOH (I) + * may also be formed. Here, we need to compare the limiting potentials of the HCOOH (I) +* pathway and subsequent pathways to determine the possible reaction pathways of the CO$_2$RR. Except for Co@MoSi$_2$N$_4$, the other four catalysts follow the HCOOH* + H$^+$ + e$^-$ → H$_2$O (I) + CHO* pathway, forming nonabsorbent CH$_4$ after progressive hydrogenation. The adsorbed O* then hydrogenates to OH*, H$_2$O*, and deadsorbed H$_2$O. All intermediates and products are adsorbed on TM atoms.

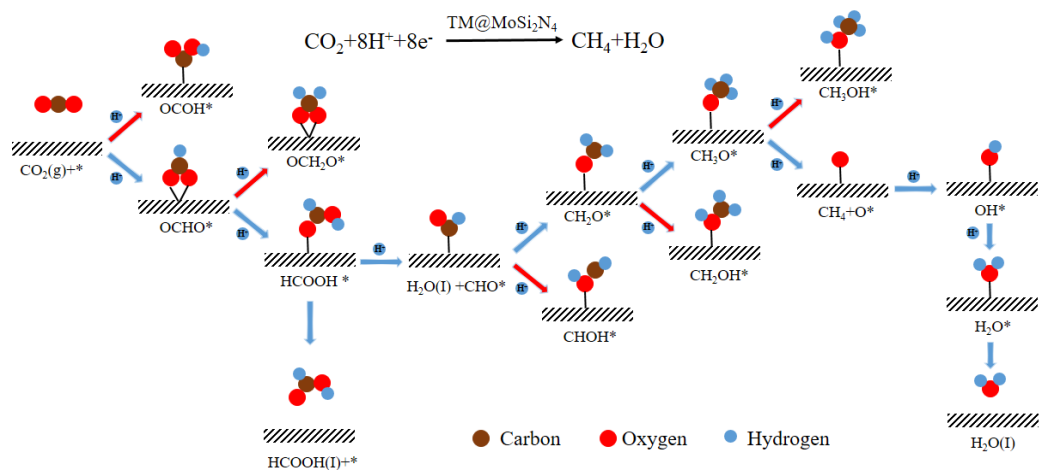

Fig. 4 Possible reaction pathways of the CO$_2$RR on TM@MoSi$_2$N$_4$, where * indicates the adsorption site on the MoSi$_2$N$_4$ surface.

The CO$_2$RR pathways to HCOOH and CH$_4$ on the TM@MoSi$_2$N$_4$ catalyst were studied in detail, as shown in Fig. 5. The overall reaction of CO$_2$RR to form CH$_4$ on the Sc@MoSi$_2$N$_4$ catalyst in the presence of hydrogen is expressed as follows:

$$CO_2 + 8H^+ + 8e^- \rightarrow CH_4 + 2H_2O \quad (1)$$

In general, the stronger the adsorption of the intermediates on the catalyst is[8], the lower the barrier for a chemical reaction that can be obtained. In Fig. 5a, we present the minimum energy pathways for CO$_2$RR to CH$_4$ on the Sc@MoSi$_2$N$_4$ catalyst.

The pathways for the formation of CH$_4$ on the Ti@MoSi$_2$N$_4$, Co@MoSi$_2$N$_4$ and Ni@MoSi$_2$N$_4$ catalysts are found to be nearly the same as those on the Sc@MoSi$_2$N$_4$ catalyst, but are thermodynamically more favorable because of the stronger interactions between Sc and the reaction intermediates. The barrier for hydrogenation of CHO* to CH$_2$O* (0.81 eV) during the formation of CH$_4$ is comparable to that for the hydrogenation of CH$_2$O* (0.74 eV). It should be noted that such a barrier is much lower than desorbed HCOOH (I) (1.83 eV). Then, CH$_2$O* is further hydrogenated to CH$_3$O* and CH$_4$* with barriers of 0.74 eV. Since the CH$_4$ product is more thermodynamically stable than HCOOH (I), the formation of CH$_4$ is likely to be the most dominant reaction pathway on the Sc@MoSi$_2$N$_4$ catalyst. For Sc@MoSi$_2$N$_4$, the rate-determining step for the formation of CH$_4$ is the hydrogenation of CHO*, with a barrier of 0.81 eV, which is the same as that for the formation of CH$_3$O* on the Ti@MoSi$_2$N$_4$ catalyst (0.83 eV, see Supplementary Fig. S8). We also investigated the pathway for the reduction of CO$_2$ on the Fe@MoSi$_2$N$_4$ and Ni@MoSi$_2$N$_4$ catalysts (see Supplementary Fig. S9, S10). We found that the interactions between Fe@MoSi$_2$N$_4$ (Ni@MoSi$_2$N$_4$) and the reactant/intermediates are slightly weak. The weak interactions between the catalyst and the reaction intermediates will lead to a large barrier for the formation of CH$_4$ on the TM@MoSi$_2$N$_4$ catalyst. For instance, the rate-determining step for the Co@MoSi$_2$N$_4$ (Ni@MoSi$_2$N$_4$) catalysts is the hydrogenation of HCOOH* (the hydrogenation of CH$_2$O*), with a barrier of 1.18 eV (1.24 eV).

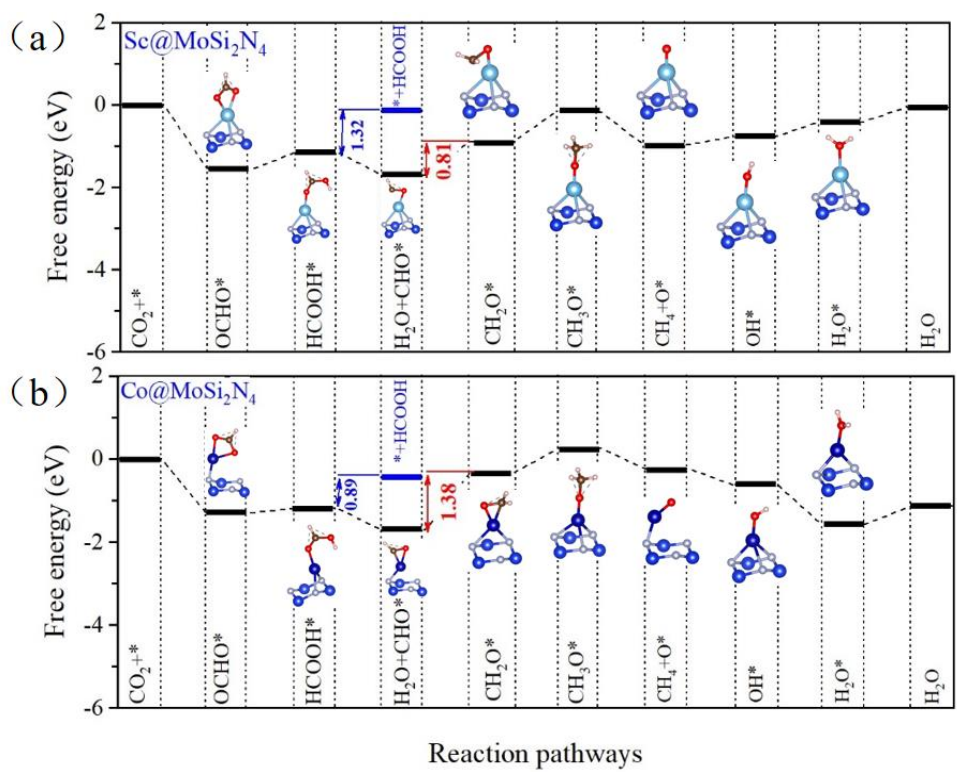

Fig. 5 CO$_2$RR paths on Sc@MoSi$_2$N$_4$ and Co@MoSi$_2$N$_4$. Free energy profile for CO$_2$ electrochemical reduction reactions along the minimum energy path at 0 V (vs. RHE) on (a) Sc@MoSi$_2$N$_4$ and (b) Co@MoSi$_2$N$_4$. The insets show the optimized configurations of the intermediates.

The formation of HCOOH follows the first two steps of the CH$_4$ pathway,

$$CO_2 + 2H^+ + 2e^- \rightarrow HCOOH \tag{2}$$

Then, HCOOH* desorbs directly from the Co@MoSi$_2$N$_4$ catalyst. The rate-determining step of the HCOOH pathway is the second step, with a barrier of 0.89 eV, which is 0.49 eV lower than the rate-determining barrier for the formation of CH$_4$ (1.38 eV). In addition, the desorption energies of HCOOH on Sc@MoSi$_2$N$_4$, Ti@MoSi$_2$N$_4$, Fe@MoSi$_2$N$_4$ and Ni@MoSi$_2$N$_4$, are 1.32 eV, 1.69 eV, 1.89 eV and 1.94 eV, respectively. Therefore, the preferred product for CO$_2$RR on the Co@MoSi$_2$N$_4$ catalyst is the HCOOH pathway rather than the CH$_4$ pathway

## 4. CONCLUSIONS

In conclusion, single metal atoms, including Sc, Ti, Fe, Co and Ni, supported on MoSi$_2$N$_4$ as electrocatalysts for CO$_2$RR were investigated by DFT calculations. A single-atom catalyst can not only maximize the efficiency of metal atoms but also exhibit excellent activity for CO$_2$RR. As evaluated by the reaction barriers, the preferred product of CO$_2$RR on the Co@MoSi$_2$N$_4$ catalyst is HCOOH with a barrier of 0.89 eV, while the Sc@MoSi$_2$N$_4$ (Ti@MoSi$_2$N$_4$, Fe@MoSi$_2$N$_4$ and Ni@MoSi$_2$N$_4$) catalyst is able to reduce CO$_2$ to CH$_4$ efficiently with a barrier of 0.81 eV (0.83 eV, 1.18 eV and 1.24 eV). These SACs based on 2D materials with thicknesses of n≥7 hold great promise for improving catalytic activity and selectivity for electrochemical the CO$_2$RR. Our study not only enriches the understanding of single atom catalysts but also provides clues for further catalyst design for CO$_2$ electroreduction in theory and experiments.

**Acknowledgements**

This work is supported by the National Natural Science Foundation of China (Grants No. 12004295). P. Li thanks China's Postdoctoral Science Foundation funded project (Grant No. 2020M673364). This work was calculated at Supercomputer Center in Suzhou University of Science and Technology.